%% file: vortex_bullets08.tex
\documentclass[pdflatex,sn-mathphys-ay]{sn-jnl}

\usepackage{graphicx}%
\usepackage{multirow}%
\usepackage{amsmath,amssymb,amsfonts}%
\usepackage{amsthm}%
\usepackage{mathrsfs}%
\usepackage[title]{appendix}%
\usepackage{xcolor}%
\usepackage{textcomp}%
\usepackage{manyfoot}%
\usepackage{booktabs}%
\usepackage{algorithm}%
\usepackage{algorithmicx}%
\usepackage{algpseudocode}%
\usepackage{listings}%

\theoremstyle{thmstyleone}%
\theoremstyle{thmstyletwo}%

\theoremstyle{thmstylethree}%

\begin{document}

\title[Three-Dimensional Optical Vortices in Nonlinear Bessel Waveguides]{Three-Dimensional Optical Vortices in Nonlinear Bessel Waveguides}


\author[1,2]{\fnm{Carlos F.} \sur{S\'{a}nchez}}
\author[2]{\fnm{Humberto} \sur{Michinel}}
\author[3,4]{\fnm{Boris}  \sur{Malomed}}
\author[2]{\fnm{\'{A}ngel} \sur{Paredes}}
\author*[2]{\fnm{Jos\'{e} R.} \sur{Salgueiro}}\email{jrs@uvigo.es}

\affil[1]{\orgdiv{Grupo de Investigaci\'on en F\'isica (GIF)}, \orgname{Universidad Polit\'ecnica
Salesiana}, \orgaddress{\city{Cuenca}, \postcode{010105}, \country{Ecuador}}}

\affil[2]{\orgdiv{Instituto de F\'isica, Computaci\'on e Ciencia Aeroespacial (IFCAE)},
\orgname{Universidade de Vigo}, \orgaddress{\street{Campus As Lagoas}, \city{Ourense},
\postcode{32004}, \country{Spain}}}

\affil[3]{\orgdiv{Department of Physical Electronics, School of Electric and Computer
Engineering, Faculty of Engineering}, \orgname{Tel Aviv University}, \orgaddress{\city{Tel Aviv}, \country{Israel}}}

\affil[4]{\orgdiv{Instituto de Alta Investigaci\'{o}n}, \orgname{Universidad de Tarapac\'{a}},
\orgaddress{\street{Casilla 7D}, \city{Arica}, \country{Chile}}}


\abstract{We explore the formation, stability, and propagation dynamics of light
bullets carrying orbital angular momentum (vorticity) in Bessel-structured
nonlinear graded-index waveguides. While these waves are prone to the
collapse in uniform three-dimensional media, we demonstrate that their
stabilization is provided by a squared-Bessel index profile in the plane
orthogonal to the propagation direction, which may play the role of the
waveguiding or antiwaveguiding potential. Different solution families are
obtained, revealing a variety of power-vs.-propagation-constant diagrams, characterizing 
the coexistence of multiple families. We analyze the propagation
dynamics in a wide experimentally accessible parameter range. Our results
contribute to the understanding of nonlinear light localization in
multidimensional settings and suggest new possibilities for using structured
light in advanced photonic applications.}



\maketitle


\section{Introduction}

Spatiotemporal optical solitons~\citep{malomed2005spatiotemporal}, often
referred to as light bullets\emph{\ }(LBs) ~\citep{silberberg1990collapse},
are photonic wave packets that maintain their self-trapped shape due to the
synergetic balance between diffraction, group-velocity dispersion (GVD), and
optical nonlinearity. Their propagation is modeled by multidimensional
nonlinear Schr\"{o}dinger equations (NLSEs)~\citep{fibichnonlinear}, which
apply to light pulses and beams in optical communications, lasers, extreme
optics, and other fields~\citep{kivshar2003optical}.

It is well known that LBs are unstable to the spontaneous onset of the
catastrophic collapse (blowup) in self--focusing media with anomalous GVD%
~\citep{bespalov1966,sulem2007nonlinear,edmundson1997unstable}. To prevent the
blowup and secure stable propagation of nonlinear optical pulses in two or
three spatial dimensions (2D or 3D), various mechanisms have been explored%
~\citep{Malomed2022},
including, \textit{inter alia}, the incorporation of higher--order terms in
the refractive index, such as saturable~\citep%
{akhmediev1992modulation,edmundson1992} or cubic--quintic nonlinearities~\citep%
{michinel2002liquid,michinel2006turning}, as well as nonlocality~\citep%
{krolikowski2001modulational,bang2002collapse,mihalache2006three,paredes2020optics}
and appropriate potential structures~\citep%
{baizakov2003multidimensional,Yu1995,Mihalache2004,Raghavan2000}. In
particular, LBs can be stabilized by means of the effective trapping
provided by waveguiding linear-index profiles~\citep%
{mihalache2005stable,diebel2016controlled,akozbek1998optical}.

Further, imparting the optical angular momentum (OAM) to optical beams is the basis
of many fundamental phenomena and applications, making the study of such beams
a research hotspot in singular optics. The loading of OAM onto the beam can be realized by 
imprinting a cross phase \citep{Liang2019,Sun2024,Sun2024a} or a winding phase onto the beam,
creating an optical vortex~\citep{nye1974dislocations,Coullet1989}.  In the case
of a self-trapped optical pulse the angular momentum leads to the formation of vortex bullets~\citep{desyatnikov2005azimuthons,desyatnikov2005optical},
whose intrinsic structure is characterized by the topological charge, alias
winding number~\citep{quiroga1997,michinel2004square,salamin2018fields}.
Despite their theoretical appeal and promising properties, the experimental
realization of stable LBs, and especially of vortex bullets, is a highly
challenging problem~\citep{hang2013ultraslow}. The control over the input beam and
securing a necessary balance between competing factors are critical for
achieving the sustained propagation of LBs and, more generally speaking,
multidimensional solitons~\citep%
{porras2019upper,chen1997self,zhang2022stabilization,Malomed2022}.

In the present paper, we explore the use of Bessel--lattice
refractive--index structures in the plane transverse to the propagation
direction~\citep{mihalache2005stable} as a means to maintain the existence and
stability of vortex bullets, cf. Refs. ~\citep%
{kartashov2005soliton,kartashov2004rotary}. For a Gaussian profile, the
existence of stable spatiotemporal vortex solitons was recently demonstrated
in the context of Bose-Einstein condensates~\citep{Sanchez2025}, which are
modeled by a variety of NLSEs in the form of Gross-Pitaevskii equations
(GPEs)~\citep{Pitaevskii2003}.
We analyze the stability of the predicted self-trapped vortex modes by dint
of the Vakhitov--Kolokolov (VK) stability criterion~\citep%
{vakhitov1973stationary,fibichnonlinear} and systematic simulations of their
propagation. Earlier, a particular case was considered~\citep{zhong2010three}
with the external potential and propagation constant specifically chosen so
that, under the Hartree approximation space-time factorization), the spatial
radial structure of the solution reduced to the Bessel functions, while the
temporal profile took the form of a hyperbolic secant. In this work, we
search for general families of solutions by addressing, in particular, the
difference between the Bessel and Gaussian waveguiding structures in axially
symmetric self--focusing optical media with anomalous GVD.

In the following section we introduce the theoretical framework and basic
equations of the model. In section~\ref{sec:numerical_solutions} we outline
the numerical scheme used in this work to produce stable vortex bullets. In
section~\ref{sec:solution_families} we present numerically generated
solution families, and in section~\ref{sec:system_stability} we analyze
their stability. The paper is concluded by section 6.

\section{The model}

\label{sec:model}

Our starting point is the NLSE for the medium with the Kerr nonlinearity\citep{kivshar2003optical} and
an axially--symmetric pattern of the linear refractive index $\Delta n\,V(r)$, where $\Delta n$ describes the scale or index contrast,

\begin{equation}
i\frac{\partial \tilde\Psi }{\partial \tilde z}+\frac{1}{2k_0}\tilde\nabla _{\perp }^{2}\tilde\Psi -\frac{\beta_2}{2}\frac{%
\partial ^{2}\tilde\Psi }{\partial \tilde{T}^{2}}+k_0\left[\Delta n V(r)+n_2 |\tilde\Psi |^{2}\right]
\tilde\Psi =0.  \label{eq:dimensional_NLSE}
\end{equation}%

This equation governs the propagation along the the $\tilde z$--direction of the
envelope $\tilde \Psi (\tilde r,\phi ,\tilde z,\tilde T)$ of a pulsed beam described by a
cylindrical coordinate system.  Parameter
$k_0=\omega_0/c$ is the vacuum vawenumber expressed in terms of
the carrier frequency $\omega_0$ and vacuum speed of light $c$. Time $\tilde T$ is expressed
in a reference system traveling with the pulse as $\tilde T=t-z/v_{g}$, where $t$ is the conventional time and $%
v_{g}$ is the group velocity. Parameter $\beta_2$ is the group velocity
dispersion (GVD) parameter and $n_2$ the second order Nonlinear (Kerr) index.
Further, the transverse Laplacian $\tilde\nabla _{\perp }\equiv
\partial _{\tilde r}^{2}+(1/\tilde r)\partial _{\tilde r}+\partial _{\phi }^{2}$ $\ $represents
the paraxial--diffraction operator. Tildes above the spatial and temporal variables and the optical
field imply they are measured in physical units. The variables are converted into the normalized form
by means of the substitution,

\begin{eqnarray}
r=k_0\sqrt{2\,\Delta n}\,\tilde r, &\qquad z= k_0\,\Delta n\,\tilde z,\nonumber \\
T=\left(\frac{2k_0\Delta n}{\beta_2}\right)^{1/2}\tilde T, &\qquad \Psi =\left(\frac{n_2}{\Delta n}\right)^{1/2} \tilde\Psi,
\label{eq:scale_eqs}
\end{eqnarray}
which leads to the following scaled equation:

\begin{equation}
i\frac{\partial \Psi }{\partial z}+\nabla _{\perp }^{2}\Psi -\sigma \frac{%
\partial ^{2}\Psi }{\partial T^{2}}+\left[ V(r)+\alpha |\Psi |^{2}\right]
\Psi =0.  \label{eq:NLSE}
\end{equation}

In this equation only two normalized parameters were kept, $\sigma$ and $\alpha$, to account respectively
for the sign of the GVD ($\sigma=\pm 1$ for normal or anomalous dispersion) and Kerr effect ($\alpha\pm1$ for  a focusing or a defocusing medium).  The spatial pattern of the axially
symmetric linear refractive index\ with strength $|\gamma|$ is defined as
\begin{equation}
V(r)=\gamma \left[ F(r)\right] ^{2}.  \label{V}
\end{equation}

This definition implies that, in the experimentally relevant setup, based on
the photorefractive material ~\citep{Efremidis2002},
the real function $F(r)$ represents the transverse structure of the
ordinary-polarization optical field illuminating the waveguide. Through the
XPM effect, $F(r)$ induces the effective potential, $\gamma \left[F(r)\right]%
^{2}$, guiding the probe beam launched in the extraordinary polarization.
Here, we consider two different transverse structures, \textit{viz}., the
Gaussian, with

\begin{equation}
F(r)=\exp (-r^{2}/4),  \label{G}
\end{equation}
and the Bessel function,
\begin{equation}
F(r)=J_{0}(r).  \label{J}
\end{equation}

In the case of the normal GVD, i.e., $\sigma =1$ in Eq.~(\ref{eq:NLSE}),
there are no solutions in the form of LBs. In this case, optical pulses do
not collapse but split, instead, after a transitional compression stage.
Note that another species of partly localized waves, previously known in the
linear context, can also be produced by the NLSE with the self-focusing
nonlinearity, \textit{viz}., the X-waves~\citep{saari1997evidence} and
X-solitons~\citep{conti2003nonlinear} (so called due to their characteristic
shape), which were also found for focusing nonlinearities.

In the case of the anomalous GVD, $\sigma =-1$, the temporal dimension can
be treated as an additional coordinate, as the full Laplacian in Eq.~(\ref%
{eq:NLSE}) is the sign-definite elliptical operator. This fact makes it
possible to produce fully localized spatiotemporal modes in the presence of
the self-focusing nonlinear response.

In the two-dimensional case (one transverse spatial dimension plus the
temporal one) there is a critical energy threshold above which the LB
initiates the collapse after passing a certain distance that depends on the
initial conditions, hence this collapse scenario is called a critical one.
In particular, the pre-collapse distance may be significantly increased by
imparting a negative chirp onto the input. If the initial energy falls below
the critical value, the input decays without starting the collapse. On the
other hand, in 3D (with the two transverse dimensions) the collapse is
supercritical, which means that the critical energy level is zero (in other
words, the input with an arbitrarily small energy may initiate the 3D
collapse) ~\citep{fibichnonlinear,sulem2007nonlinear,Zakharov2012}.

In the present work, we consider the case of the self-focusing ($\alpha =1$)
in the anomalous-GVD regime, $\sigma =-1$, aiming to obtain vortex-bullet
stationary solutions, introduced as

\begin{equation}
\Psi (r,\phi ,z,T)=\psi (r,T)e^{i\ell \phi }e^{i(\beta z-\omega t)},
\label{eq:stationary_state}
\end{equation}%
where $\psi $ is a real function, $\beta $ is a real propagation constant,
and $\ell $ is the integer vorticity, alias the topological charge. We focus
on fundamental vortices, with $\ell =1$. For a given value of $\beta $ the
solution is uniquely determined by its total energy, i.e., in the scaled
units, the 3D norm $N$ of the wavefunction (\ref{eq:stationary_state}),
which is defined by Eq. (\ref{eq:energy}) below.

\section{Numerical solutions}

\label{sec:numerical_solutions}

Substituting ansatz~(\ref{eq:stationary_state}) in Eq.~(\ref{eq:NLSE}), one
arrives at the eigenvalue problem based on the following equation for the
real field $\psi (r,T)$:

\begin{equation}
-\beta \psi +\frac{\partial ^{2}\psi }{\partial T^{2}}+\frac{\partial
^{2}\psi }{\partial r^{2}}+\frac{1}{r}\frac{\partial \psi }{\partial r}-%
\frac{\ell ^{2}}{r^{2}}\psi +[V(r)+\psi ^{2}]\psi =0.  \label{eq:beta}
\end{equation}%
This equation was solved numerically applying the finite-differences scheme
(for variables $r$ and $T$) and solving iteratively the resulting nonlinear
algebraic problem~\citep{salgueiro2007computation}.
The presence of the first derivative in Eq.~(\ref{eq:beta}) breaks the
symmetry of the linear equation system to be solved as a part of
each Newton's iteration. Because solvers intended for nonsymmetric matrices
are much less efficient than those for symmetric ones, we apply the known
transformation $\psi (r)=r^{-1/2}u(r)$, which eliminates the first radial
derivative, leading to the following equation:

\begin{equation}
\frac{\partial ^{2}u}{\partial T^{2}}+\frac{\partial ^{2}u}{\partial r^{2}}-%
\frac{(\ell ^{2}-1/4)}{r^{2}}u+\left[ V(r)-\beta +\frac{u^{2}}{r}\right] u=0,
\label{eq:final}
\end{equation}%
with the 3D norm

\begin{equation}
N=2\pi \int_{-\infty }^{+\infty }\int_{0}^{\infty }|\psi |^{2}rdr\,dT = 2\pi
\int_{-\infty }^{+\infty }\int_{0}^{\infty }|u|^{2}dr\,dT.  \label{eq:energy}
\end{equation}

\begin{figure}[tbph]
\begin{center}
\includegraphics[width=\columnwidth]{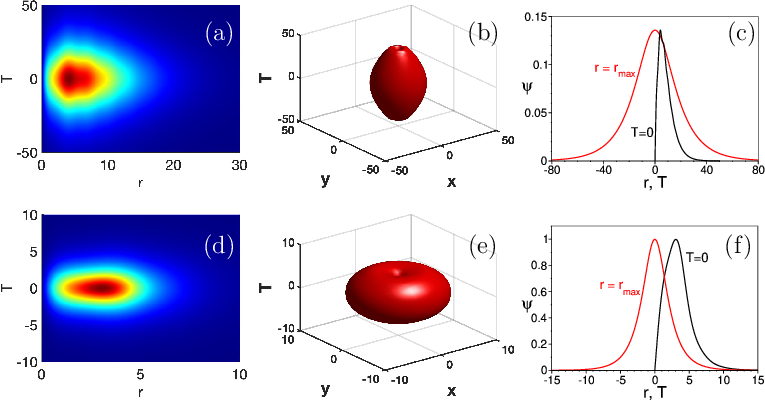}
\end{center}
\caption{Two examples of numerical solutions of Eq.~(\protect\ref{eq:beta})
for the Bessel lattice defined as per Eqs.~(\protect\ref{V}) and (\protect
\ref{J}), with $\ell =1$. Top row correspond to a solution with parameters
$\protect\gamma =2$ and $\protect\beta =0.05$. Shown are views of the contour in (a)
of the field in the plane $\left( r,T\right) $, and the 3D isosurface view in (b). Panel (d)
shows the the spatial and temporal profiles, plotted in cross-sections $T=0$ and $%
r=r_{\max }$, where $r_{\max }$ is the radial coordinate corresponding to
the maximum of $|\protect\psi |$. Botom row shows contour (d), isosurface (e) and
profiles (f) for another solution with parameters $\protect\gamma =2$ %
and $\protect\beta =0.25$.}
\label{fig:fig1}
\end{figure}

Solutions to Eq.~(\ref{eq:final}) must satisfy the boundary condition $%
u(r=0)=0$, hence producing a regular solution for $\psi (r)$ at $%
r\rightarrow 0$ in spite of the diverging factor $r^{-1/2}$ in the above
transformation.

Examples of numerically found solutions are displayed in Fig.~\ref{fig:fig1}
for the waveguide with the squared-Bessel potential
profile defined as per Eqs.~(\ref{V}) \ and (\ref{J}). The vortex structure,
with the cylindrical symmetry, demonstrates itself in the spatial
transversal plane $(x,y)$ [alias ($r,\phi $), in the polar coordinates].
Thus, the stationary states can be adequately represented in terms of the
spatial and temporal coordinates, $r$ and $T$.

\begin{figure}[th]
\begin{center}
\includegraphics[width=\columnwidth]{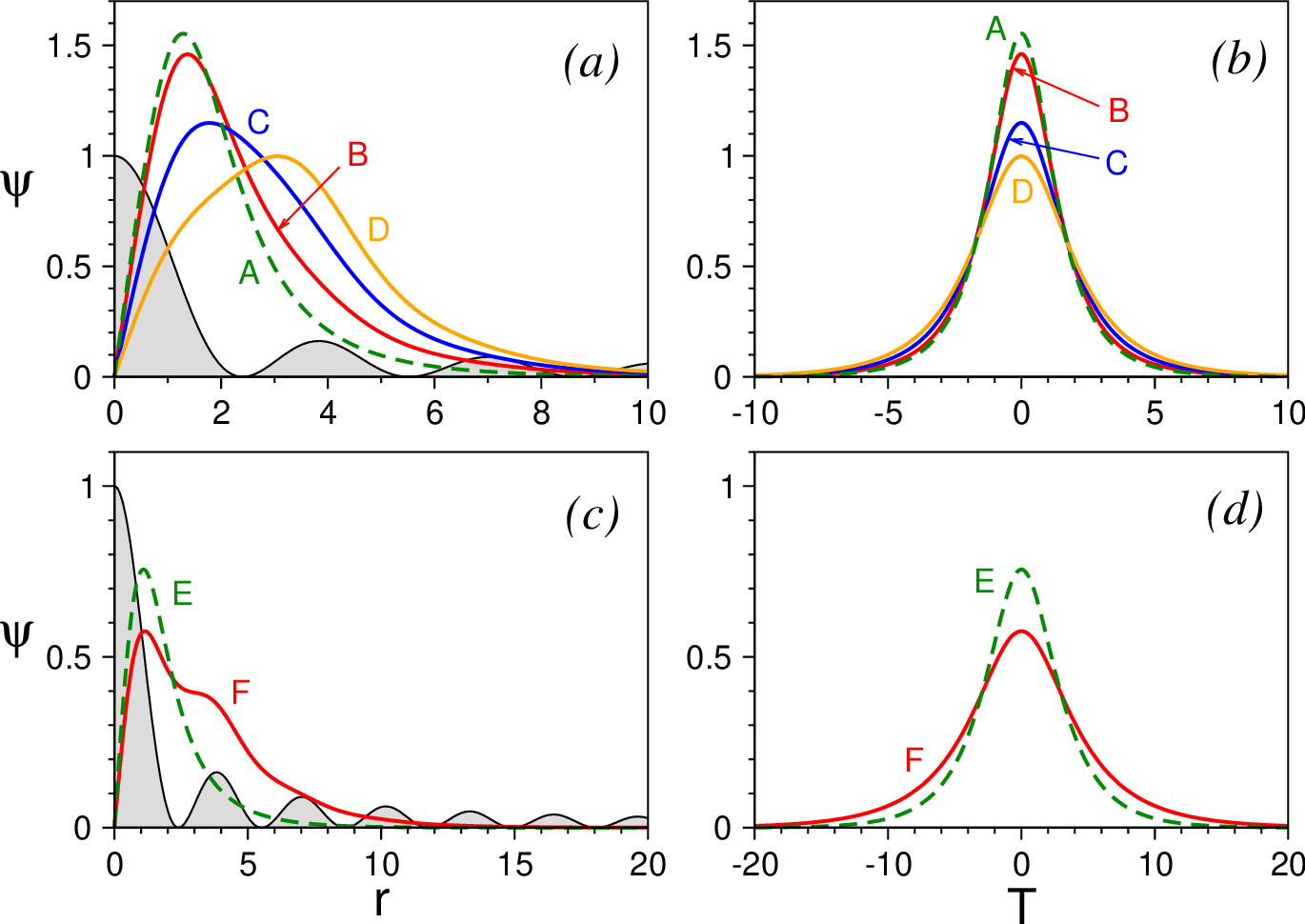}
\end{center}
\caption{Examples of the numerically calculated profiles of the vortex
bullets along $r$ and $T$ (the left and right panels, respectively) for $%
\protect\beta =0.25$ and $\protect\gamma =2$ in (a,b), and $\protect\gamma =4
$ in (c,d). The thin black line bounding the shadowed area is the normalized
squared-Bessel profile, $J_{0}^{2}(r)$. The colored solid and dashed curves
correspond, severally, to the Bessel and Gaussian waveguiding structures.
The profiles labelled with letters correspond to the white dots in Figs.~%
\protect\ref{fig:fig6} and \protect\ref{fig:fig8} marked with the same labels. }
\label{fig:fig3}
\end{figure}

To give a realistic estimate of related physical quantities, we recall that a
typical index change in the photorefractive induced guide is 
$\Delta n\sim6\times10^{-4}$~\citep{Efremidis2002}. The dispersion,
depending on the material and waveguide geometry, typically
features the zero-dispersion point in the near infrared, similar to the
usual optical fibers. Around such a point, for telecom wavelengths of interest,
the dispersion is measured in tenths of ps/(nm$\cdot$ km). Fixing this value
as $D=10\,$ps/(nm$\cdot$ km) in the anomalous-dispersion region, the respective
second-order dispersion parameter is obtained as
$\beta_2=-\lambda^2 D/(2\pi c)=-1.27\times10^{-14}$\,s$^2$/m. Then, for
a typical wavelength, $\lambda=1.55\,\mu$m, one can use Eqs.~(\ref{eq:scale_eqs})
to evaluate physical values which correspond to units on the axes in Fig.~\ref{fig:fig3}. Thus,
for spatial coordinate and temporal coordinate we obtain, respectively, $\tilde{r}/r = 7.13\,\mu$m
and $\tilde{T}/T = 1.62$\,ns.

In Fig.~\ref{fig:fig3} we additionally plot several radial and temporal
profiles of the stationary states for the propagation constant $\beta =0.25$%
, corresponding to two different waveguides with strength $\gamma =2$ (a-b)
and $\gamma =4$ (c-d). Solid lines are the modal profiles obtained for the
squared-Bessel waveguiding structure, indicated by the shadowed area in the
plots, and the dashed curves correspond to the Gaussian structure defined as
per Eqs.~(\ref{V}) and (\ref{G}). Both structures are chosen so as to have
the same width.

As it can be seen in the plots, for $\gamma =2$ and propagation constant $%
\beta =0.25$, there is only one stationary state or three different ones,
maintained by the Gaussian or Bessel waveguides respectively. This is due
to the fact that the $J_{0}^{2}(r)$ function has adjacent maxima that can
dramatically change the guiding properties of the system through the
interplay with the self-focusing nonlinearity. For $\gamma =4$ (deeper
waveguides) there is a single stationary state with $\beta =0.25$, in the
Gaussian and Bessel waveguides alike. Naturally, the profiles with higher
energy (larger values of norm $N$) are more focused in the temporal
dimension, as seen in the right-column plots of Fig.~\ref{fig:fig3}. Note
also that, for deeper waveguides [larger $\gamma $ in Eq.~(\ref{V})],
smaller energy is required to induce the nonlinear self-trapping.
Accordingly, positions of maxima of the stationary states in stronger
waveguides are closer to local maxima of the refractive index.

\begin{figure}[th]
\begin{center}
\includegraphics[width=\columnwidth]{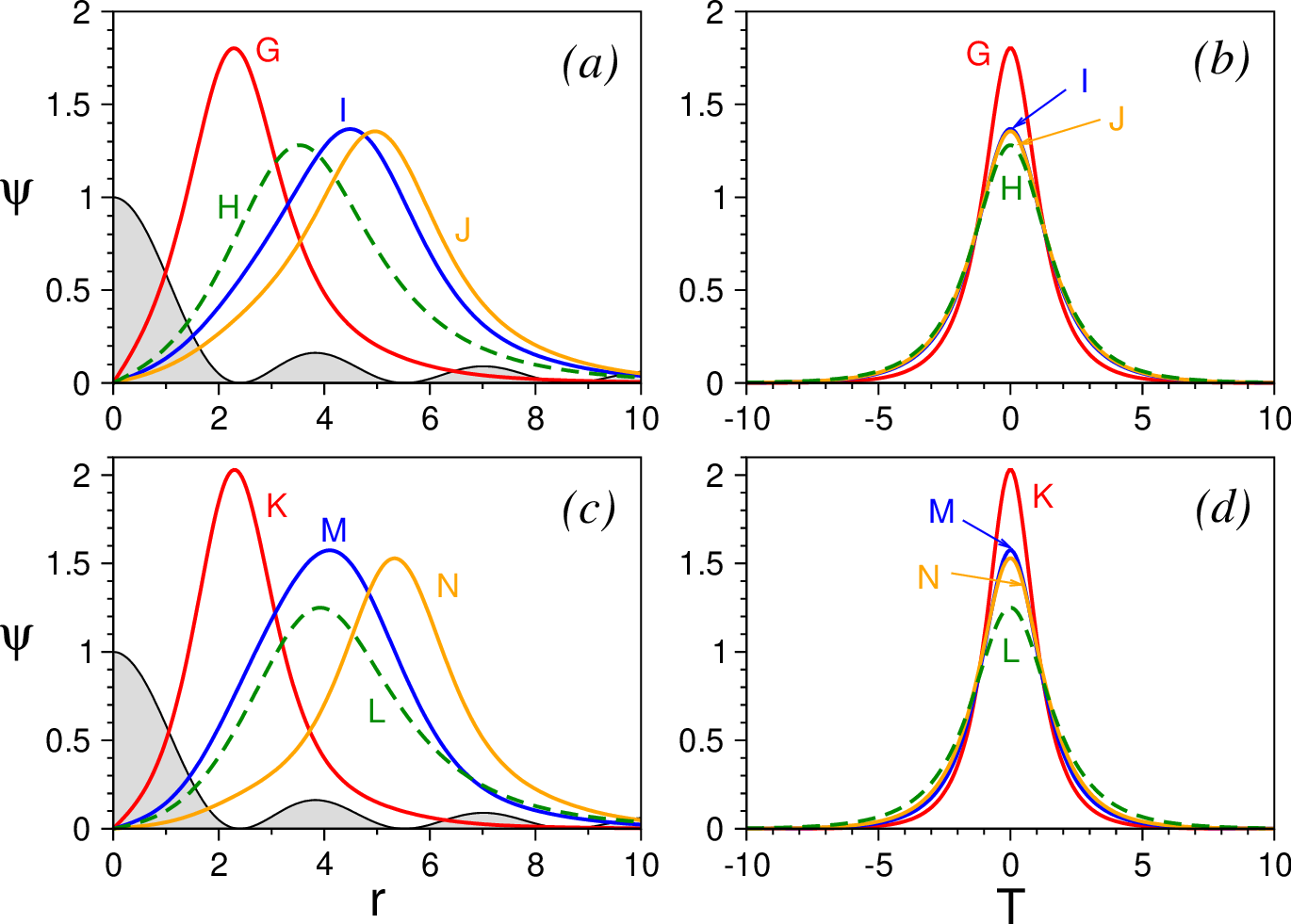}
\end{center}
\caption{The same as in Fig.~\protect\ref{fig:fig3}, but for the AWG, with $%
\protect\gamma <0$ and $\protect\beta =0.2$. Top and bottom panels
correspond to $\protect\gamma =-2$ and $\protect\gamma =-4$ respectively.
The profiles labelled with letters correspond to the white dots marked with the
same labels in Figs.~\protect\ref{fig:fig7} and \protect\ref{fig:fig8}. }
\label{fig:fig4}
\end{figure}

Figure~\ref{fig:fig4} is similar to Fig.~\ref{fig:fig3} but for the\textit{\
antiwaveguides} (AWGs) ~\citep{Gisin1995,Bortman2001,Kaplan2002},
with $\gamma =-2$ or $\gamma=-4$ (the top and bottom rows respectively) and $\beta
=0.25$. Due to the self--focusing effect, stationary LBs can be found in
this case too. As in the case presented in Fig.~\ref{fig:fig3}, there are
several eigenstates with the same value of $\beta $ for the Bessel AWG and
a single one for the anti-Gaussian. Notably, both for the lower ($\gamma =-2$%
) and higher ($\gamma =-4$) Bessel AWGs, three LBs with different radial
profiles are found. Note that the temporal shapes of the LBs are nearly
identical for close values of the energy ($N$), as the temporal shape is not
directly affected by the transverse (anti)waveguiding structure.

In Fig.~\ref{fig:fig4} the situation for the Gaussian AWG agrees with what
might be expected: for the stronger AWG structure (larger $-\gamma $), more
energy is necessary to achieve the self-trapping, the maxima being located
at almost the same position for $\gamma =-2$ and $\gamma=-4$. On the other
hand, in the case of the Bessel AWG with its set of local potential minima,
a more complex interplay with the self-focusing is observed. For instance,
comparing two red profiles in the left-column panels, one can see that the
trapping may occur around the same position, with less energy for the higher
AWG structure, which might not be expected in the absence of the local
potential minima (as in the Gaussian case).

\begin{figure}[th]
\begin{center}
\includegraphics[width=\columnwidth]{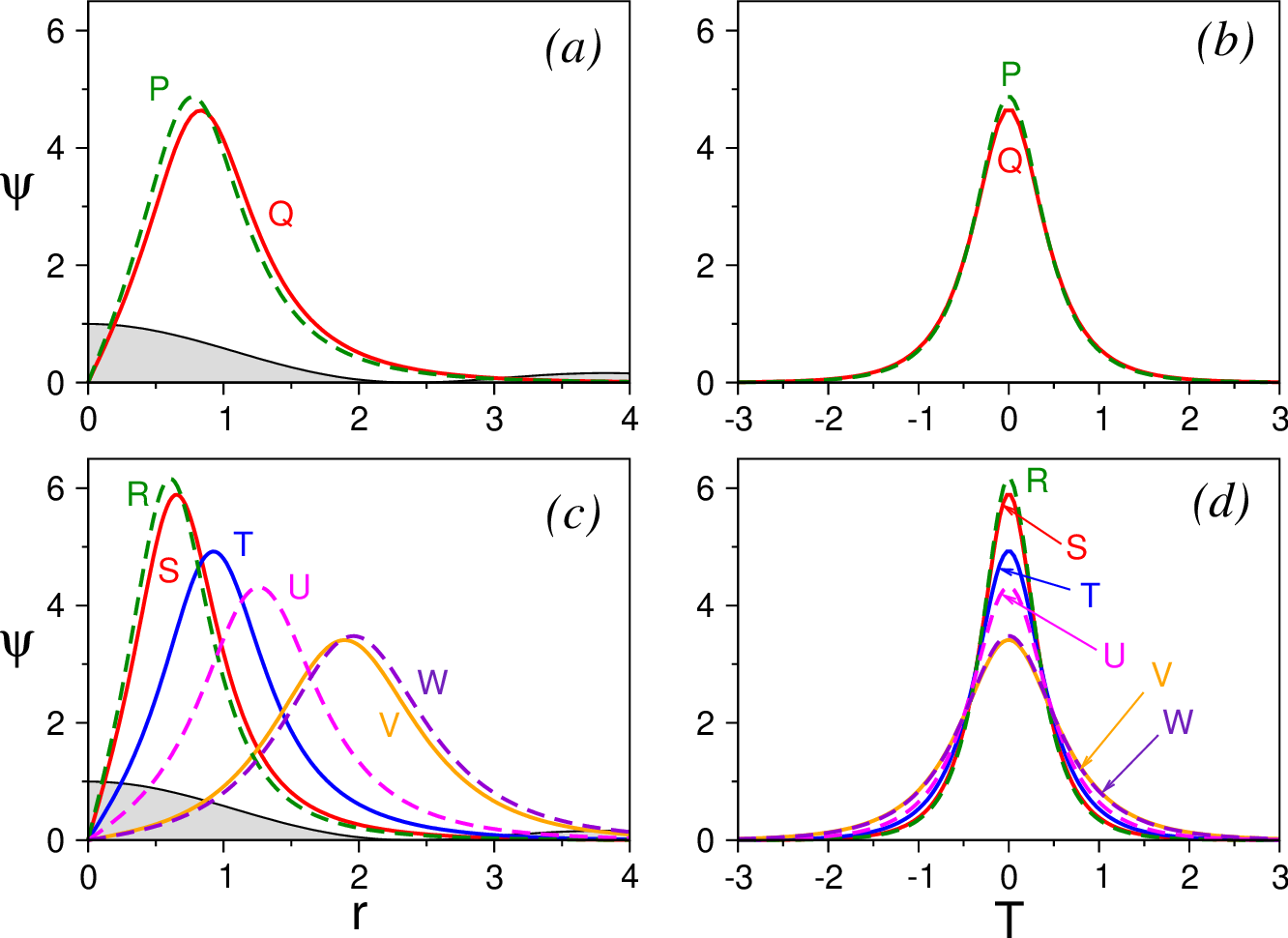}
\end{center}
\caption{The same as in Fig.~\protect\ref{fig:fig4}, but for $\protect\beta %
=1.6$. The top and bottom panels correspond to $\protect\gamma =-2$ and $%
\protect\gamma =-4$, respectively. The profiles labelled with letters
correspond to the white dots marked with the same labels in Figs.~\protect\ref%
{fig:fig7} and \protect\ref{fig:fig8}. }
\label{fig:fig5}
\end{figure}

Figure~\ref{fig:fig5} is similar to Fig.~\ref{fig:fig4} but for $\beta =1.6$,
with the top and bottom panels corresponding to $\gamma =-2$ and $\gamma=-4
$, respectively. Contrary to Fig.~\ref{fig:fig4}, Fig.~\ref{fig:fig5}
demonstrates several eigenstates with the same value of $\beta $, both for
the Bessel and Gaussian AWGs with $\gamma=-4$.
On the other hand, for the weaker AWG with $\gamma =-2$, a single
stationary state with $\beta =1.6$ is found for the Gaussian and Bessel
structures alike. As above, the temporal distributions are nearly identical
for close values of the pulse energy. It is also seen that both the radial
and temporal profiles are quite similar for the Bessel and Gaussian
structure, which is not surprising as the self-trapping is chiefly
determined by the nonlinearity.

\section{Solution families}

\label{sec:solution_families}

\begin{figure}[th]
\begin{center}
\includegraphics[width=\columnwidth]{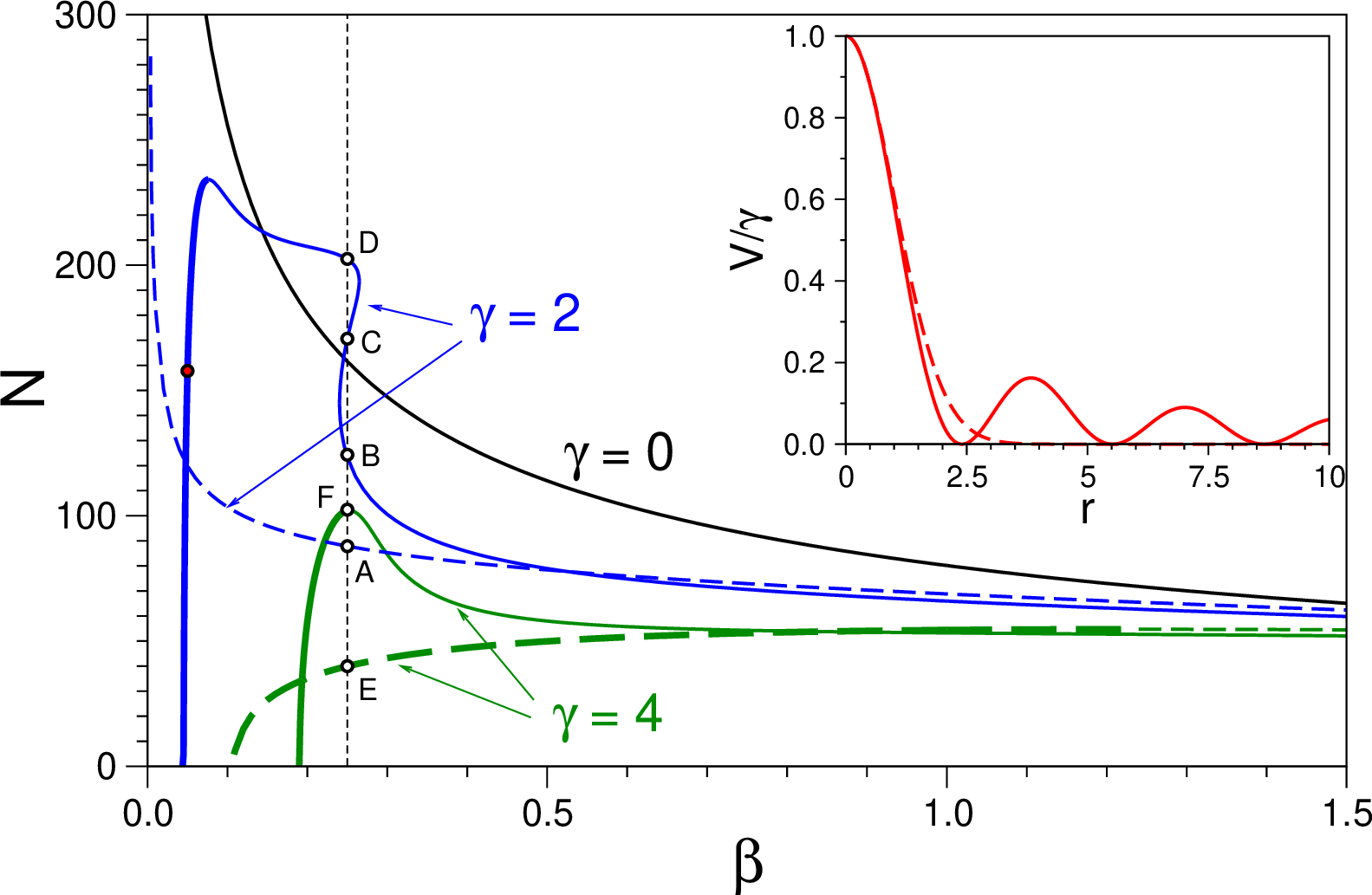}
\end{center}
\caption{Norm $N$ of several families of stationary LB solutions vs. the
propagation constant $\protect\beta $ for the values of the waveguide
strength $\protect\gamma =0$, $2$, and $4$. Continuous and dashed curves
correspond, respectively, to the LBs maintained by the Bessel and Gaussian
waveguiding structures. Thicker sections of the curves indicate the domain
where LBs are stable. The black curve with $\protect\gamma =0$ is included
for the comparison. White empty dots correspond to the LB profiles of the
same label plotted in Fig.~\protect\ref{fig:fig3}. The filled dot indicates
the stable LB in Fig.~\protect\ref{fig:fig1}. The inset shows the respective
waveguiding profiles, \textit{viz}., the squared-Bessel and Gaussian ones
(continuous and dashed curve, respectively).}
\label{fig:fig6}
\end{figure}

To gain insight into the physical properties of the LB stationary states, we
classify them into families parametrized by the values of their propagation
constant and energy (i.e., norm $N$). In Fig.~\ref{fig:fig6} we present the
families for the guiding profile ($\gamma >0$), by means of $N(\beta )$
curves, for $\gamma =2$ and $\gamma =4$. The case of $\gamma =0$ corresponds
to the uniform Kerr medium, and is also shown in the plot. The continuous
and dashed curves correspond, respectively, to the LBs maintained by the
squared-Bessel and Gaussian waveguiding structures. The inset displays both
index profiles. A filled dot indicates the stationary
solution from Fig.~\ref{fig:fig1}, and a vertical gray line at $\beta =0.25$
cuts the curves at empty white spots that correspond to the stationary
solutions shown in Fig.~\ref{fig:fig3}.

As seen in Fig.~\ref{fig:fig6}, in the absence of the waveguiding structure (%
$\gamma =0$) the norm diverges at $\beta \rightarrow 0$, while it is well
known that no stable solutions exist in this case ~\citep%
{fibichnonlinear,Malomed2022}. The presence of the waveguiding structure ($%
\gamma >0 $) dramatically changes the behavior, generating a cutoff value
for the propagation constant in the linear limit ($N=0$), which corresponds
to the guiding supported exclusively by the refractive-index distribution.
As shown by the solid curves with $\gamma =2$ and $\gamma =4$, the energy
initially grows with $\beta $, attaining a maximum and subsequently
decreasing. The strength threshold to observe this cutoff is lower for the
squared-Bessel profile than for the Gaussian one as is evident
from the dashed curve correspondent to
$\gamma=2$, which does not shown either the cutoff point or the initial raising
energy. We again stress that multiple states with the same value of the
propagation constant $\beta$ can be maintained by the Bessel structure with
$\gamma =2$, as it was mentioned above, in the relation to Fig.~\ref%
{fig:fig3}.

\begin{figure}[th]
\begin{center}
\includegraphics[width=\columnwidth]{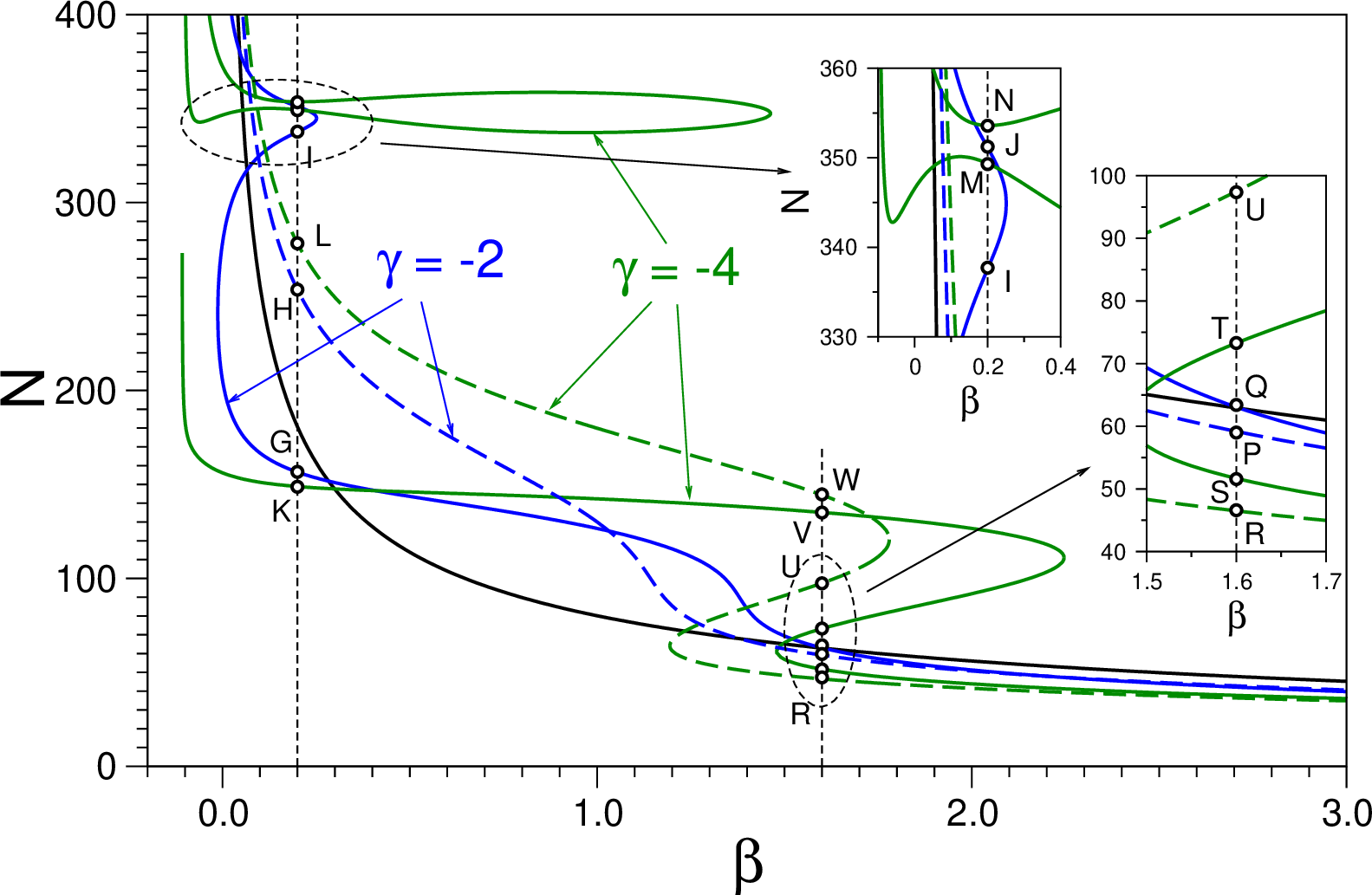}
\end{center}
\caption{The same as Fig.~\protect\ref{fig:fig6} for different values $%
\protect\gamma <0$ of the AWG strength parameter. White dots labelled with
letters correspond to the LB profiles of the same label displayed in Figs.~%
\protect\ref{fig:fig4} and \protect\ref{fig:fig5}. Insets reproduce zoomed
regions of the main plot for a better clarity.}
\label{fig:fig7}
\end{figure}

In Fig.~\ref{fig:fig7} we show the families for the AWG profile ($\gamma <0$%
). In this case, $N$ may increase indefinitely with decreasing values of $%
\beta$. Also, valid values of $\beta$, i.e. those leading to bounded modes,
may now be negative due to the AWG potential.
As the potential strength, $-\gamma $, increases, there appears S-shaped
features on the $N(\beta )$ curve, which is demonstrated by Fig.~\ref%
{fig:fig7} for $\beta =-4$ in the interval of, roughly speaking, $1.5<\beta
<2.3$. On the other hand, at large values of $N$ ($N\sim 350$) the same
figure demonstrates a prolate loop, detached from the main curve. Large
values of ring radius of the corresponding states suggest that they are
related to the first maximum of the respective Bessel potential.

\begin{figure}[th]
\begin{center}
\includegraphics[width=\columnwidth]{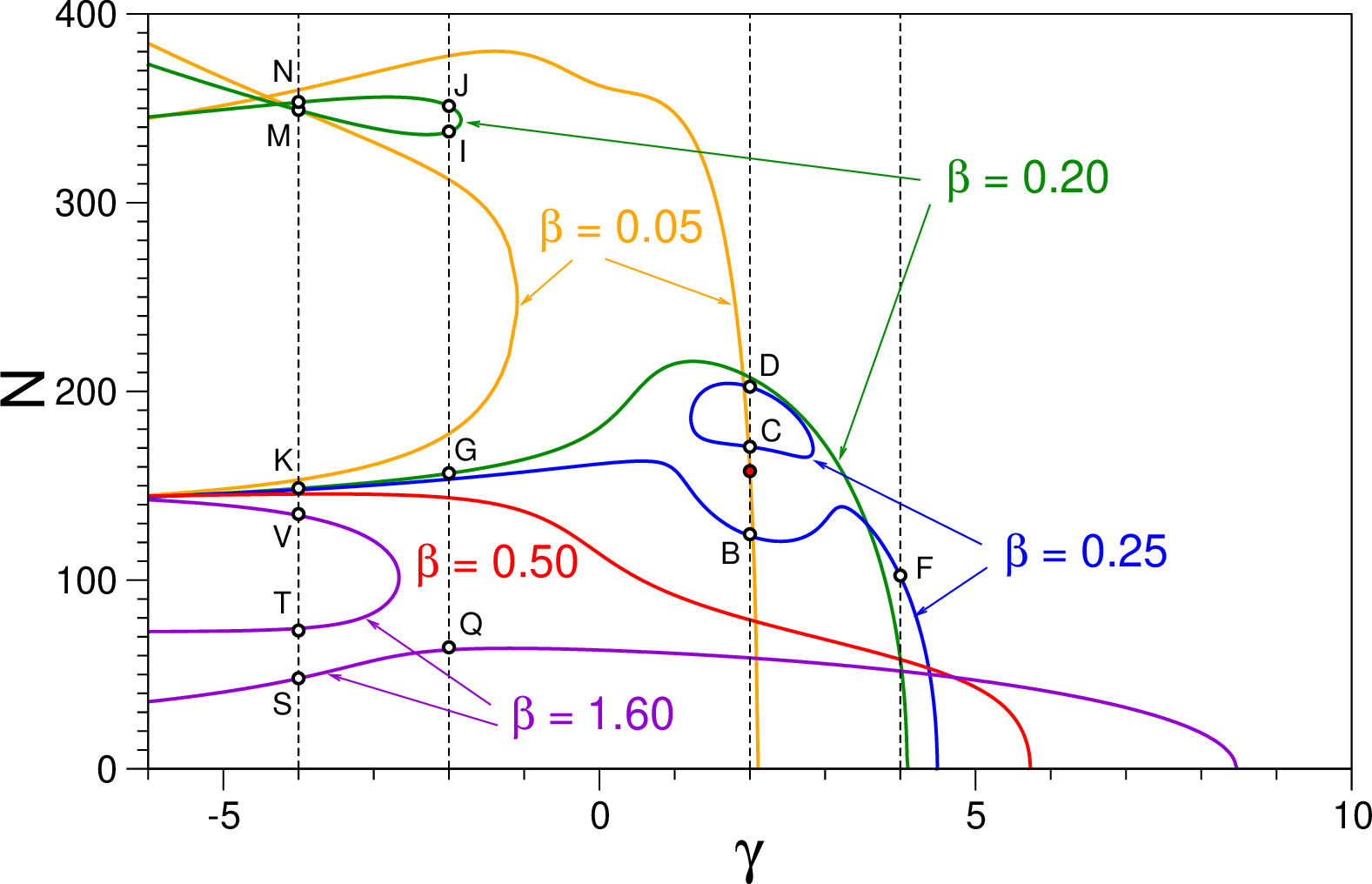}
\end{center}
\caption{The norm $N$ of the stationary LB vs. the Bessel-potential strength $%
\protect\gamma $, for different fixed values of the propagation constant $%
\protect\beta $. The red dot indicates the stable state shown in Fig.~%
\protect\ref{fig:fig1}. White dots with labels correspond to the profiles
with the same label displayed in Figs.~\protect\ref{fig:fig3}, \protect\ref%
{fig:fig4} and \protect\ref{fig:fig5}.}
\label{fig:fig8}
\end{figure}

In Fig.~\ref{fig:fig8} we plot the norm $N$ vs. strength $\gamma $ of the
squared-Bessel potential for different values of the propagation constant, $%
\beta $. The plot helps to identify intervals of $\gamma $ where multiple
stationary solutions exist. They feature a loop shape -- for instance, the
loop for $\beta =0.25$ in the interval $1.22<\gamma <2.84$, where three
different stationary solutions are found. At particular values of $\gamma $
the loop self-crosses, as it happens for $\beta =0.20$, close to $\gamma =-4$
(in the case of the AWG structure), implying the detachment of the loop from
the main curve. For a particular value of $\beta $, the curve attains $N=0$
(e.g., at $\gamma \approx 8.5$ for $\gamma =1.60$ in Fig.~\ref{fig:fig8})
which implies reaching the limit where amplitude decreases to zero as the
temporal-width diverges to infinity. This happens as the nonlinearity is no
longer able to keep the state bounded in the temporal direction, although
the potential, if it is strong enough, would confine the state in the
spatial one.


\section{The stability analysis}

\label{sec:system_stability}

\subsection{The Vakhitov-Kolokolov (VK) criterion}

The stability of the stationary solutions can be assessed, in the
quasi-analytical form, by means of the VK criterion which states that a
necessary condition for the stability of a stationary solution, taken as per
Eq.~(\ref{eq:stationary_state}), is $dN/d\beta >0$ ~\citep%
{vakhitov1973stationary,fibichnonlinear}. Conversely, the solution is
definitely unstable in the case of $dN/d\beta <0$. The VK criterion actually
implies the absence of unstable eigenmodes of small perturbations with
purely real instability growth rates. Note that the ubiquitous instability
of vortex solitons against spontaneous splitting into fragments is not
detected by the VK criterion because the splitting instability is accounted
by complex growth rates. Eventually, numerical simulation of the propagation
should verify which stationary states are truly stable.

In the free-space case, with $\gamma =0$, the monotonously decreasing $%
N(\beta )$ dependence in Fig.~\ref{fig:fig6} predicts, as expected, the
fully unstable LB\ family. Analyzing the other curves in Fig.~\ref{fig:fig6}
with respect to the VK criterion we arrive at the following conclusions:

\begin{itemize}
\item For $\gamma =2$, the dashed curve for the Gaussian potential remains
monotonically decreasing, which means the full instability.

\item For $\gamma =4$, the green dashed curve for the Gaussian potential
monotonously increases at $\beta <1.24$, attaining a maximum, and decreases
at $\beta >1.24$. Hence stable stationary states may be possible at $\beta
<1.24$. These possibilities for the Gaussian trapping potential have been
analyzed in Ref.~~\citep{Sanchez2025} in the context of Bose-Einstein
condensates and it was concluded they were fully stable.

\item For $\gamma =2$ with the squared-Bessel potential the situation is
more complex, suggesting regions of multi-stability and alternating stable
and unstable fragments with stationary solutions having the same value of
the propagation constant, corresponding to different values of norm $N$,
around the region with $\beta =0.25$.

\item For $\gamma =4$ with the squared-Bessel potential the situation is
qualitatively similar to that for the Gaussian potential, with the value of $%
\beta =0.26$ marking the transition from stable to unstable solutions.
\end{itemize}

The physical implication of these observations is that the effective Bessel
potential introduces a rich stability/instability pattern and the
corresponding bifurcation structure. Indeed, the $N(\beta )$ curves exhibit
loops and turning points, reflecting the impact of the potential on the
existence and stability of the LBs. The conclusion is that, while
low--energy solutions may be stabilized by the potential, the stability is
maintained in specific intervals of $\beta$.

With respect to Fig.~\ref{fig:fig7}, that corresponds to the AWG potential
with $\gamma <0$, the conclusions are:

\begin{itemize}
\item For $\gamma =-2$, the dashed curve for the Gaussian potential is
monotonically decreasing, indicating the full instability in this case.

\item For $\gamma =-4$, the dashed curve for the Gaussian case displays a $Z$%
-like shape in the interval of $1.2<\beta <1.8$, which is compatible with
possible bistability.

\item For $\gamma =-2$ with the squared-Bessel potential (the continuous
curve), three stationary states with different values of the norm exist in
the interval of $0<\beta <0.25$, and only a single state at $\beta >0.25$.
According to the VK criterion, stable states may exist for increasing $%
N(\beta )$ branches. 

\item For $\gamma =-4$ with the squared-Bessel potential (the continuous
line), the situation is still more complex, due to conspicuous oscillations
of the $N(\beta )$ curve, indicating possible multi-stability windows, that
should be explored by means of numerical simulations.
\end{itemize}

To conclude, stationary states which are potentially stable according to the
VK criterion may exist both for the waveguiding and AWG potentials, for both
the squared-Gaussian and Bessel potentials. The depth of the waveguide plays
a crucial role in predicting the VK-stability regions.

\subsection{Numerical simulations.}

\begin{figure}[th]
\begin{center}
\includegraphics[width=\columnwidth]{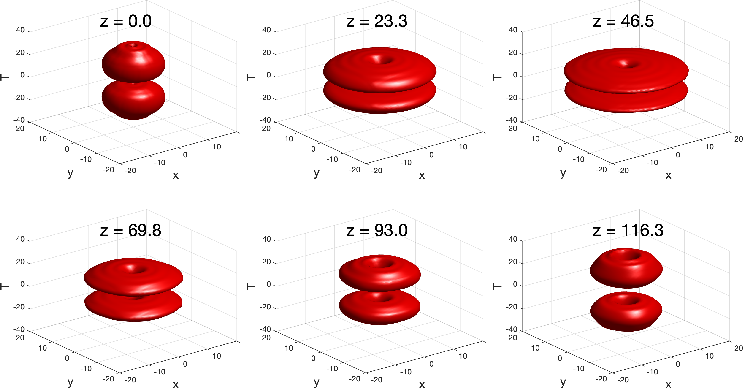}
\end{center}
\caption{An example of the elastic collision (rebound) of two stable LBs
with opposite signs (the phase difference of $\protect\pi $), for $\protect%
\gamma =4$, $\protect\beta =0.21$. Top row: the approach stage; bottom row: the
bouncing stage. }
\label{fig:fig9}
\end{figure}

To finally verify the stability of the stationary LBs, we used the standard
beam-propagation method, based on the finite-difference discretization of
Eq.~(\ref{eq:NLSE}). The results are reported in Figs.~\ref{fig:fig9}-\ref%
{fig:fig11}.

As concerns the LBs in the guiding-type potential ($\gamma >0$), we have
found that they are indeed stable in the region of $\partial N/\partial\beta
>0$, as predicted by the VK criterion. We have checked this conclusion by
means of the long-distance propagation, up to $z=500$.

We also simulated collisions between LBs, as shown in Fig.~\ref{fig:fig9},
which demonstrate that such collisions are quasi-elastic. In this case, the
colliding LBs are out of phase, so that they bounce back after the
collision. In case both LBs are in phase they temporary mix to subsequently
emerge after crossing themselves.


\begin{figure}[th]		
\begin{center}
\includegraphics[width=\columnwidth]{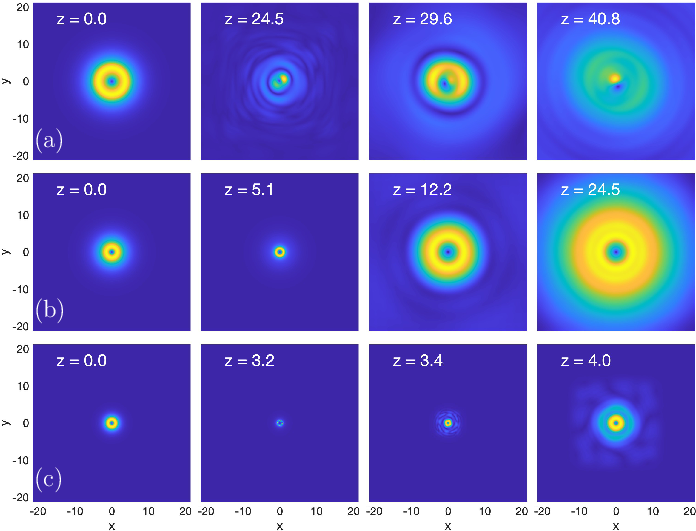}
\end{center}
\caption{Examples of the collapsing evolution of VK-unstable LBs, guided by
the squared-Bessel potential with $\protect\gamma =2$, see Eqs.~(\protect\ref%
{V}) and (\protect\ref{J}): (a) for $\protect\beta =0.2$, (b) for $\protect%
\beta =0.25$, (c) for $\protect\beta =1.0$.}
\label{fig:fig10}
\end{figure}

In the direct simulations the VK-unstable LBs, belonging to the region of $%
dN/d\beta <0$, collapse to a small spot and then spread out. The collapsing
distance is shorter for larger $\beta $, as seen in the examples shown in
Fig.~\ref{fig:fig10}. Interestingly, the above-mentioned splitting
(azimuthal) instability is not directly observed in this regime with $\gamma
>0$. It may occur only at the post-collapse stage.

\begin{figure}[th]
\begin{center}
\includegraphics[width=\columnwidth]{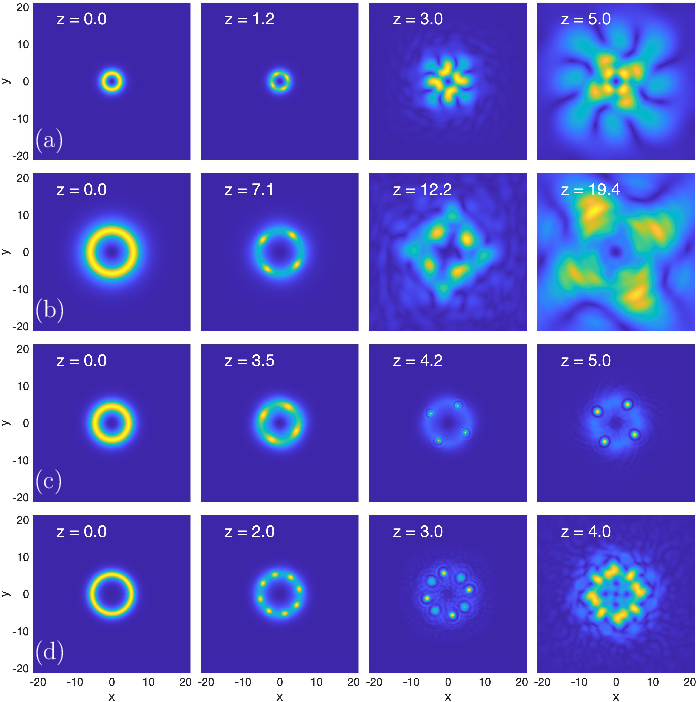}
\end{center}
\caption{Instability scenarios of different LBs for the AWG
squared-Bessel-like potential, with $\protect\gamma =-4$. (a) $\protect\beta %
=1.0$, (b) $\protect\beta =0.05$, (c) $\protect\beta =0.5$ (this case
belongs to the loop of the $N(\protect\beta )$ curve, see Fig. \protect\ref%
{fig:fig7}), (d) $\protect\beta =1.2$ (this case also belongs to the loop). }
\label{fig:fig11}
\end{figure}

In the case of an AWG potential, we observe three instability stages in the
evolution: an initial azimuthal splitting (producing four fragments in most
cases), followed by the collapse of the fragments into thin spikes, and then
spreading of the spikes which ultimately fuse into a broadening ring. This
instability scenario takes place for all values of $\beta $ (see examples
for $\gamma =-4$ and different values of $\beta $ in Figs.~\ref{fig:fig11}%
(a-b), including those in the regions of Z-shaped $N(\beta )$ curves),
revealing that in this case meeting the VK criterion does not lead to
stability. In the upper loop of the curve (the one detached from the main
one), see Fig. \ref{fig:fig7}, the instability scenario is similar, see
Fig.~\ref{fig:fig11}(c), but at large enough values of $\beta $, where the
radius of the vortex ring is larger, splitting instability produces eight
fragments instead of four, as shown in Fig.~\ref{fig:fig11}(d).

\section{Conclusion}

In this work, we have studied the formation, stability, and propagation
dynamics of LBs (light bullets) carrying orbital angular momentum
(vorticity) in nonlinear graded-index waveguides. The analysis was performed
for the fundamental vortex modes with topological charge $\ell =1$. We have
focused on two profiles of the effective potential in the plane orthogonal
to the propagation direction, \textit{viz}., the Gaussian and the
squared-Bessel ones, which may be waveguiding or AWG (antiwaveguiding). Both
potentials can be experimentally realized in optics, using photorefractive
materials. The crucially important stability problem for the vortex LBs was
tackled with the help of the VK (Vakhitov-Kolokolov) criterion; then, the
stability was accurately identified by means of systematic propagation
simulations. We have thus demonstrated the existence of stable vortex LBs
under experimentally accessible conditions. We have also extensively
analyzed the evolution of unstable LBs, which displays various dynamical
phenomena.

These results contribute to the understanding of the nonlinear light
localization in multidimensional systems and suggest new possibilities for
controlling structured light in advanced photonic applications.

\bmhead{Acknowledgements}

This work is a part of the R\&D\&i projects PID2023-146884NB-I00 funded by Ministerio de Ciencia e Innovación of Spain and GPC ED431B 2024/42 funded by Xunta de Galicia (Spain)

\section*{Declarations}

\begin{itemize}
\item Funding. This research was supported by projects PID2023-146884NB-I00 of Ministerio de Ciencia e Innovación of Spain and GPC ED431B 2024/42 of Xunta de Galicia (Spain)

\item Conflict of interest/Competing interests. The authors have no relevant financial or non-financial interests to disclose.

\item Data availability. The datasets generated during the current study and selected to compose the different figures in the manuscript are available in \url{https://doi.org/10.5281/zenodo.18861294}

\item Author contribution. All authors contributed to the research conception and preparation of this manuscript and have approved the final version.\end{itemize}




\input bibliography.bbl

\end{document}